\documentclass{ws06-procs}
\begin{document}
\title{Steps toward a classifier for the Virtual Observatory.
I. Classifying the SDSS photometric archive.}
\author{D'Abrusco Raffaele$^1$}
\address{Department of Physical Sciences, University of Napoli Federico II, via Cinthia 9, 80126 Napoli, ITALY.}
\author{Staiano Antonino}
\address{Department of Physical Sciences, University of Napoli Federico II, via Cinthia 9, 80126 Napoli, ITALY.}
\author{Prof. Longo Giuseppe}
\address{INFN - Napoli Unit, Dept. of Physical Sciences, via Cinthia 9, 80126, Napoli, ITALY.\\
         Department of Physical Sciences, University of Napoli Federico II, via Cinthia 9, 80126 Napoli, ITALY.} 
\author{Paolillo Maurizio}
\address{Dept. Physical Sciences, University 'Federico II', Naples, Italy  \& INFN -
         Napoli Unit, Dept. of Physical Sciences, via Cinthia 9, 80126,
         Napoli, ITALY} 
\author{De Filippis Elisabetta}
\address{Dept. Physical Sciences, University 'Federico II',
          Naples, Italy \& INAF-Italian National Institute of Astrophysics, via del Parco
          Mellini, Rome, I }
         
\maketitle

\abstracts{Modern photometric multiband digital surveys produce
large amounts of data that, in order to be effectively exploited,
need automatic tools capable to extract from photometric data an
objective classification.\\ We present here a new method for
classifying objects in large multi\-parametric photometric data
bases, consisting of a combination of a clustering algorithm and a
cluster agglomeration tool. The generalization capabilities and
the potentialities of this approach are tested against the
complexity of the Sloan Digital Sky Survey archive, for which an
example of application is reported.}

\section{Introduction}
In the last few years the astronomical community is experiencing a
tremendous growth in the size, quality and accessibility of
databases. This trend will accelerate in the coming years due to the advent
of large dedicated survey telescopes and to the
implementation of the International Virtual Observatory infrastructure. 
There is the practical fact that the extraction of useful
information from such datasets cannot be effectively performed with
traditional tools, and from the methodological point of
view, the wealth of information contained in such huge data sets
imposes to abandon old conceptual schemes largely based on the 3-D
visualization capability of human minds and to adopt "ad hoc"
statistical pattern recognition, classification and visualization
methods. The application of such algorithms to the astronomical case is all
but trivial, due to the complexity of astronomical data which
usually present strong non linear correlations among parameters
and are highly degenerate. Among the other tools, especially relevant
to the astronomical case are those which deal with the identification and
visualization of groups of objects sharing the same physical
properties.

\section{The methods}

The method outlined here follows a hierarchical approach which,
starting from a preliminary clustering performed using a
clustering algorithm, the "Probabilistic Principal Surfaces",
followed by a second phase that uses the Negative Entropy concept
and a dendrogram structure to agglomerate the clusters found in
the first phase.

\subsection{PPS - Principal Probabilistic Surfaces}

Probabilistic Principal Surfaces (PPS)
\cite{chang_2000} are a nonlinear extension of
principal components, in that each node on the PPS is the average
of all data points that projects near/onto it. PPS define a
non-linear, parametric mapping $\mathbf{y}(\mathbf{x};
\mathbf{W})$ from a $Q$-dimensional latent space $(\mathbf{x} \in
\mathbb{R}^Q )$ to a $D$-dimensional data space $(\mathbf{t} \in
\mathbb{R}^D)$, where normally $Q << D$. The function
$\mathbf{y}(\mathbf{x}; \mathbf{W})$ (defined continuous and
differentiable) maps every point in the latent space to a point
into the data space. Since the latent space is $Q$-dimensional, 
these points will be confined to a $Q$-dimensional manifold non-linearly embedded into
the $D$-dimensional data space. In our
method, the points belonging to the parameter space will be
projected on the surface of 2-dimensional sphere. The
visualization capabilities of the PPS can prove very useful in
several aspects of the data interpretation phase such as, for
instance, the localization of data points lying far away from the
more dense areas (outlayers), or of those lying in the overlapping
regions between clusters, or to identify data points for which a
specific latent variable is responsible.

\subsection{NEC - Negentropy Clustering}

Most unsupervised methods require the number of clusters to be
provided \emph{a priori}, a serious
problem when exploring large complex data sets where the number of
clusters can be very high or unpredictable. A simple
treshold criterium is not satisfactory in most astronomical
applications due to the high degeneracy and the noisiness of the
data which lead to erroneous agglomeration, while a
different approach based on the combination of a similarity
criterium based on the concept of Negative Entropy and the use of
a dendrogram as agglomerative algorithm is achievable. We
implement the Fisher's linear discriminant which is a
classification method that first projects high-dimensional data
onto a line, and then performs a classification in the projected
one-dimensional space \cite{bishop_1995}. 
On the other hand, we define the differential entropy H of
a random vector ${\mathbf y} = (y_1, \ldots, y_n)^T$ with density
$f(.)$ as $ H({\mathbf y}) = \int f({\mathbf y}) \log f({\mathbf
y}) d{\mathbf y}$ so that negentropy $J$ can be defined as $
J(\mathbf{y}) = J({\mathbf y}_{Gauss}) - H(\mathbf{y})$, where
$\mathbf{y}_{Gauss}$ is a Gaussian random vector of the same
covariance matrix as $\mathbf{y}$.  The Negentropy can be interpreted as a measure of non-Gaussianity and, since it is invariant for invertible linear transformations, it is obvious
that finding an invertible transformation that minimizes the
mutual information is roughly equivalent to finding directions in
which the Negentropy is maximized. Our
implementation of the method use approximations of Negentropy that
give a very good compromise between the properties of the two
classic non-Gaussianity measures given by kurtosis and Negentropy.
Negentropy can be used to agglomerate with an unsupervised method
the clusters (regions) found by the PPS approach. The only {\it a
priori} information is a dissimilarity threshold $T$. We suppose
to have $c$ multi-dimensional regions $X_i$ with $i=1,\ldots,c$
that have been defined by the PPS approach, then passing these
regions to the Negentropy Clustering algorithms which, in
practice, measures whether two clusters could or could not be
modeled by one single Gaussian or, in other words, if the two
regions can be considered to be aligned or as part of a greater
data set.

\section{The data}

All data used in this work are extracted from the Data Release 4
of the Sloan Digital Sky Survey (\cite{SDSS}). This spectroscopic subsample will
constitute the ''knowledge base'' on which we have founded the
labelling of the unsupervised ones. The SDSS also provides, for
each object in the SpS, a spectroscopic classification index
called \emph{specClass}. All objects are classified in
\emph{specClass} as either a quasar, high-redshift quasar (with $z
> 2.2$ ), galaxy, star, late-type star, or unknown (ranging in
value of \emph{specClass} from 0 to 6) by matching emission lines
found in the observed spectrum against a list of common galaxies
and quasar emission lines. We have extracted from the SDSS-4
spectroscopic subsample a catalog containing $\sim 600000$
objects, excluding from the query only the objects labelled as 'SKY'
according to \emph{specClass}. The percentage distribution of the
resulting sample respect to the \emph{specClass} index is the
following: \emph{specClass} 0:  1.5\%  , \emph{specClass} 1: 8.5 \%,
\emph{specClass} 2: 78.7\%, \emph{specClass} 3: 8\%, \emph{specClass} 4: 0.1\% ,
\emph{specClass} 6: 2.8\%. Furthermore we excluded from this sample
drawn from the spectroscopical SDSS-DR4 data all objects fainter
than $r$=18, thus obtaining $\sim$ 43000 records.

\section{The experiment}

The unsupervised clustering method here presented is based on the
combined use of PPS and NEC algorithms. We first applied the PPS
algorithm to the sample of spectroscopically selected SDSS DR-4
objects using as parameters for the clustering the 4 colors
obtained from model magnitudes (u-g,g-r,r-i,i-z) of SDSS archive.
We fixed the number of latent variables and latent bases of the
PPS to 614 and 51 respectively, so obtaining at the end of this
step 614 clusters, each formed by objects which only respond to a
certain latent variables. We chose a large number of latent
variables in order to obtain an accurate separation of objects and
to avoid that any group of distinct but near points in the
parameter space could be projected in the same cluster by chance.
The clusters so found by PPS algorithm are graphically represented
by groups of points with the same color (a different color for
each cluster) on the surface of a 2-d sphere embedded in the
3-dimensional latent space. These groups of
objects are then input to the totally unsupervised agglomeration
NEC algorithm, whose only free parameter is the dissimilarity
threshold $T$, as above mentioned. We performed a \emph{plateau}
analysis to determine the optimal value of this threshold: we
performed different experiments with $T$ varying over a wide
range, then selected the central value of intervals of $T$ for
which the number of final clusters is constant. The number of
clusters resulting from the NEC aggregation is 31. We present in
table (\ref{tabella}) a collection of the most interesting of
these clusters after labelling each object with its \emph{specClass} index.

\begin{table}[h]
\tbl{\emph{specClass} distribution of clusters found by NEC
algorithm after PPS initialization.\label{tabella}}
{\begin{tabular}{@{} c c c c c c c c @{}}
\toprule Cl. n    & SP0  & SP1& SP2& SP3& SP4& SP6 \\
1 &69 &145 &9362 &48 &0 &12\\
2 &25 &133 &13370 &10 &0 &12\\
3 &149 &132 &63 &64 &0 &5\\
4 &44 &3396 &1530 &189 &67 &1\\
5 &202 &85 &447 &2428 &6 &10\\
6 &26 &125 &13728 &12 &0 &12\\
7 &0 &0 &0 &0 &0 &484\\
8 &1 &1 &1 &0 &0 &329\\
9 &541 &1507 &127 &4750 &18 &1\\
10 &89 &474 &2117 &19 &4 &529\\
\end{tabular}}
\end{table}

\section{Conclusions}
As can be seen from table (\ref{tabella}), different groups of clusters, according to their \emph{specClass} composition, are found. There is a significant fraction of clusters  dominated by just one kind of \emph{specClass} objects and contaminated by few objects flagged with different values of \emph{specClass}. Otherwise, other clusters show a quite homogeneous mixture of spectral type objects, with a prominence of stars(\emph{specClass}=1,6) - quasars(\emph{specClass}=3,4) and galaxy(\emph{specClass}=2) - unknown(\emph{specClass}=0) mixtures. Only few clusters show comparable proportions of all objects. A more profound analysis of these mixed clusters and the comparison between the colours and other photometric and spectroscopic informations for the same objects will hopefully cast light upon the associations between objects with different values of \emph{specClass}, and remove the degeneracy in the colour space.


\begin{thebibliography}{3}

\bibitem{chang_2000} K. Chang and J. Ghosh, {\it IEEE Transactions on 
Pattern Analysis and Machine Intelligence.}, {\bf 23(1)}: 22-41, (2000).

\bibitem{bishop_1995} C. M. Bishop, {\it Neural Networks for Pattern Recognition},
Oxford University Press, (1995).

\bibitem{SDSS} J. K. Adelman-McCarthy et al., {\it Astrophys. J. Suppl.}, 
{\bf 162}, 38-48, (2006).

\end{thebibliography}
\end{document}